# Spin-orbit interaction of light induced by transverse spin angular momentum engineering

Zengkai Shao[1*], Jiangbo Zhu[2*], Yujie Chen[1], Yanfeng Zhang[1†], and Siyuan Yu[1,2†]

[1]School of Electronics and Information Engineering, State Key Laboratory of Optoelectronic Materials and Technologies, Sun Yat-sen University, Guangzhou 510275, China.

[2]Photonics Group, Merchant Venturers School of Engineering, University of Bristol, Bristol BS8 1UB, UK.

[*]These authors contributed equally to this work.

[†]email: zhangyf33@mail.sysu.edu.cn; s.yu@bristol.ac.uk

We report the first demonstration of a direct interaction between the extraordinary transverse spin angular momentum in evanescent waves and the intrinsic orbital angular momentum in optical vortex beams. By tapping the evanescent wave of in a whispering-gallery-mode-based optical vortex emitter and engineering the transverse-spin state carried therein, a conversion between the transverse-spin angular momentum and the intrinsic orbital angular momentum carried by the emitted vortex beam takes place. This unconventional interplay between the spin and orbital angular momenta allows the regulation the spin-orbital angular momentum states of the emitted vortex. In the reverse process, it further gives rise to an enhanced spin-direction coupling effect in which waveguide or surface modes are unidirectionally excited by an incident optical vortex, with the directionality jointly controlled by the spin and orbital angular momenta states of the vortex. The identification of this previously unknown pathway between the polarization and spatial degrees of freedom of light enriches the spin-orbit interaction phenomena, and can enable a variety of functionalities employing spin and orbital angular momenta of light in applications such as communications and quantum information processing.

Light waves possess intrinsic spin and orbital angular momentum (SAM and OAM), as determined by the polarization and spatial degrees of freedom of light[1-3]. These two components are separately observable in paraxial beams[4-7], whereas it is well known that fundamentally such a distinction faces difficulties in light fields with high nonparaxiality and/or inhomogeneity[8-11]. In fact, spin-orbit interactions (SOI) can be widely observed in light through scattering or focusing[12,13], propagation in anisotropic/inhomogeneous media[14,15], reflection/refraction at optical interfaces[16,17], etc. Notably, the spatial and polarization properties of light are coupled and SOI phenomena must be considered in modern optics dealing with sub-wavelength scale systems, including nano-photonics and plasmonics[18-22]. A variety of novel functionalities utilizing structured light and materials are underpinned by SOI of light, e.g., optical micro-manipulations[23], high-resolution microscopy[24], and beam shaping with planar structures (metasurfaces)[25].

On the other hand, the study in SOI over the past few years is accompanied by a rising interest in the transverse spin angular momentum of light, which has been revealed by recent advances in optics as a new member in the optical angular momentum (AM) family[26-29]. In sharp contrast to the longitudinal SAM predicted by Poynting[1], the transverse SAM exhibits spin axis orthogonal to the propagation of light[28,30]. Transverse SAM can be typically found in highly inhomogeneous light fields, including surface plasmon polaritons[26], evanescent waves of guided and un-guided modes[22,28] and strongly focused beams[31], where longitudinal field components emerge due to the transversality of electromagnetic waves[32]. Light fields possessing transverse SAM can enable various applications in bio-sensing, nano-photonics, etc. More interestingly, transverse spin in evanescent waves also originates from the SOI in laterally confined propagating modes[11], or can also be interpreted as the quantum spin Hall effect (QSHE) of light[33-35], and thus giving rise to robust spin-controlled unidirectional coupling at optical interfaces[18,21,22,36]. This extraordinary characteristic of transverse SAM results in the breaking of the directional symmetry in mode excitation at any interface supporting evanescent waves, and can find applications in optical diodes[37], chiral spin networks[38,39], etc.

The ability to simultaneously tailor light fields in the polarization and spatial degrees of freedom via SOI phenomena has allowed for new functionalities in structured light manipulations[40]. Furthermore, combing SOI and transverse SAM control will provide a more versatile platform for processing of light fields in the full AM domain. In this paper, we present an enrichment of the SOI effects revealed by the engineering of transverse spin in evanescent waves. Our method evolves from an optical vortex emitter based on a planar integrated whispering-gallery mode (WGM) resonator, which emits beams with precisely controllable total angular momentum (TAM)[41,42]. Here we demonstrate that the engineering of transverse spin in the evanescent waves of WGMs in the resonator leads to the spin-to-orbital AM conversion in the emitted beams. This is the first demonstration of an SOI effect that features the interaction between the transverse SAM and intrinsic OAM of light, providing a promising pathway towards more sophisticated light manipulation via SOI phenomena. By reversing the emission process, we further demonstrate directional coupling of optical vortices into this integrated photonic circuitry, with the direction of the waveguide modes jointly controlled by the spin and orbital AM states, realizing the selective reception of vector vortices without separate polarization and spatial phase manipulation. These results can be used to bring novel functionalities to nano-photonic devices, e.g., encoding and retrieving photonic states in the SAM-OAM space, and provide the guidelines for the design of nano-photonic chiral interface between travelling and bounded vector vortices.

## Results

**Transverse spin in optical vortex emitter.** The schematic of the platform for the investigation of transverse spin engineering based SOI is shown in Figure 1a, where a single-transverse-mode ring resonator is coupled with a two-port access waveguide and embedded with periodic angular scatterers in the inner-sidewall evanescent region of the waveguide. With the sub-wavelength scatterers arranged in a second-order grating fashion, the diffracted first-order light from the evanescent fields of WGMs collectively produce a vortex beam carrying optical OAM and travelling perpendicular to the resonator plane[41]. In addition, the emitted vortex beams exhibit cylindrically symmetric polarization and intensity distributions, and thus referred to as cylindrical vector vortices (CVVs)[42,43].

Generally, for the quasi-transverse-electric (TE) WGMs propagating in the high-index waveguide, a local longitudinal electric component ($E_\varphi$) exists in the sidewall evanescent waves and is in quadrature phase with respect to the radial component ($E_r$) (see Figure 1b), as a direct result of the strong lateral confinement and transversality condition[32]. Consequently, the local SAM in the evanescent field exhibits a 'transverse' spinning axis in the $z$ direction[45], being orthogonal to the local propagation direction ($+\varphi$ or $-\varphi$) of the WGM. Note that for quasi-TE WGMs, the transverse SAM at the inner- and outer- sidewalls always has opposite spin directions, and the transverse spin can also be flipped by injecting light from the alternative ports 1 or 2 and exciting counter-clockwise (CCW) or clockwise (CW) WGMs, as shown in Figure 1b.

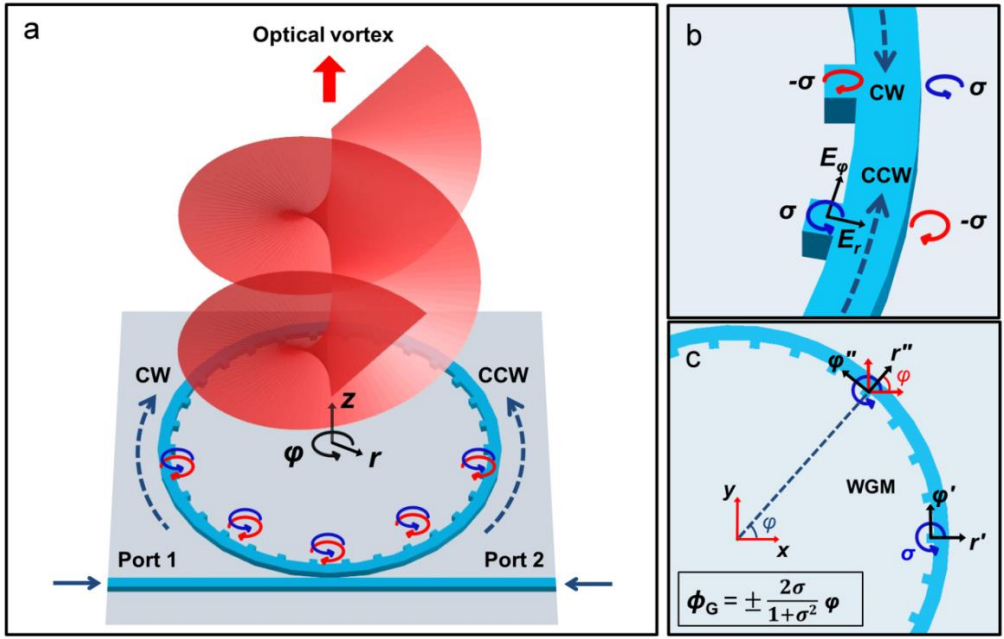

**Figure 1**. Illustration of the concepts. (a) Schematic of the platform for the investigation of transverse spin induced SOI effect. A single-transverse-mode ring resonator is coupled with an access waveguide and embedded with sub-wavelength scatterers arranged as 2$^{nd}$ order grating in the evanescent wave region. (b) Each WGM possesses transverse spin of opposite signs in the inner- and outer-resonator evanescent waves, and clock-wise (CW) and counter clock-wise (CCW) WGMs present opposite transverse spins on each side of the resonator. (c) Illustration of the transverse-spin-dependent geometric phase acquired by the vector evanescent wave as the WGM travels around the resonator. For CCW and CW WGMs, a rotation angle of $\mp \varphi \, \mathbf{z}$ is experienced by the local coordinates to be aligned with the global reference frame (i.e., from ($r''$, $\varphi''$) to ($x$, $y$)) for phase comparison of different locations, and the geometric phase acquired by evanescent wave is $\Phi_G = \pm 2\sigma/(1+\sigma^2) \, \varphi$.

**Interaction of transverse-spin and OAM.** The emission of CVVs from such structures can be generally described in the form of transfer matrices as $\mathbf{E}_{out} = \mathbf{M}_2 \mathbf{M}_1 \mathbf{E}_{in}$. By assuming the WGM evanescent wave maintains a uniform distribution around the resonator, the generic input light for the matrices is the inner sidewall evanescent wave and can be written in the locally transverse and longitudinal polarization basis. Here the CCW propagating WGM is considered as an example and thus $\mathbf{E}_{in} \propto e^{ip\varphi}[E_r \, E_\varphi]^T$ (see Supplementary Note 1 for details), where the integer $p > 0$ is the azimuthal mode number and $E_z$ is negligible at the sidewalls[46].

Firstly, the perturbation to WGM evanescent waves induced by the scatterers is expressed by the matrix

$$\mathbf{M}_1 = \begin{bmatrix} W_1 & 0 \\ 0 & W_2 \end{bmatrix} \cdot e^{i\delta(\varphi)} \qquad (1)$$

where $\delta(\varphi) = -q\varphi$ (see supplementary material of ref. 41) is the azimuthal phase acquired by the second-order grating scattering, $q$ is the number of scatterers, and $W_i$ ($i = 1, 2$) is a coefficient quantifying the scatterers' modulation on the field strength of the electric components. Here we define the transverse-spin state in the perturbed evanescent wave $|\mathbf{M}_1| \cdot \mathbf{E}_{in} \propto e^{ip\varphi}[W_1 E_r \, W_2 E_\varphi]^T$ based on the ratio of the two cylindrical components as[28]

$$\sigma = \begin{cases} \dfrac{W_1 E_r}{i W_2 E_\varphi}, & |W_1 E_r| \le |W_2 E_\varphi| \\[6pt] \dfrac{i W_2 E_\varphi}{W_1 E_r}, & |W_1 E_r| > |W_2 E_\varphi| \end{cases} \qquad (2)$$

$\sigma$ ($|\sigma| <= 1$) is a real number as $E_\varphi$ and $E_r$ always oscillate in quadrature with each other at the sidewalls[46], and it directly characterizes the (spatial) transverse-spin density in the evanescent wave as $S_\perp \propto \sigma$ (refs 6,11). For the transverse SAM of left (right) handed spin here, $\sigma > 0$ ($< 0$). In addition, the vector fields of WGMs travelling along the resonator experience a rotation of local coordinate frame, which is described by the matrix

$$\mathbf{M}_2 = \begin{bmatrix} \cos\varphi & -\sin\varphi \\ \sin\varphi & \cos\varphi \end{bmatrix} \qquad (3)$$

By applying the transfer matrices $\mathbf{M}_1$ and $\mathbf{M}_2$, the Jones vector of the output CVV becomes

$$\mathbf{E}_{out} \propto \frac{1+\sigma}{2\sqrt{1+\sigma^2}} e^{i(l_{TC}-1)\varphi} \begin{bmatrix} 1 \\ i \end{bmatrix} \mp \frac{1-\sigma}{2\sqrt{1+\sigma^2}} e^{i(l_{TC}+1)\varphi} \begin{bmatrix} 1 \\ -i \end{bmatrix} \qquad (4)$$

where $l_{TC} = p - q$ is defined as the topological charge (TC)[41]. Here the Jones vector is formulated in the global reference frame with the x- and y-polarization basis (i.e., $[E_x\ E_y]^T$). The constituent left- ($[1\ i]^T$) and right-hand ($[1\ -i]^T$) circular polarized (CP) vortices are out-of- and in-phase, respectively, when following the two definitions in Equation (2). It is straightforward to find that the CVV possesses the SAM and OAM components per photon as (see Supplementary Note 2)

$$S_z = \frac{2\sigma}{1+\sigma^2}\hbar, \quad L_z = \left(l_{TC} - \frac{2\sigma}{1+\sigma^2}\right)\hbar \qquad (5)$$

where $\hbar$ is the reduced Planck constant. Note that $S_z$ here, which should be distinguished from the spatial transverse spin density $S_\perp$, is the SAM in CVVs averaged over the transverse x-y plane. More profoundly, the variation in magnitude from the local density ($S_\perp \propto \sigma$) to the average SAM ($S_z \propto 2\sigma/(1+\sigma^2)$) is associated with a transverse-spin dependent geometric phase that stems from the rotation of local vector field. To be more specific, the Pancharatnam phase[44,47] is used to described the spatial phase variation in the CVVs of space-variant polarization state (see Supplementary Note 3)

$$\Phi_P = l_{TC}\varphi + \frac{2\sigma}{1+\sigma^2}\varphi \qquad (6)$$

where $l_{TC}\varphi = p\varphi - \delta(\varphi)$ is the scattering phase solely resulted from the first-order diffraction of grating[41]. Meanwhile, the second term, $\Phi_G = 2\sigma\varphi/(1+\sigma^2)$, has a pure geometric nature and arises from the rotation of local transverse-spin state while WGMs travel around the resonator (see Figure 1c). It should be emphasized that $\Phi_G$ differs from all the previously discussed geometric phases of light that can be identified either in artificial anisotropic structures[44] or light beams of curvilinear trajectories[15], and originates essentially from the coupling between the transverse SAM of guided light and the rotation of light's path. Nevertheless, this transverse-spin dependent geometric phase is still in accordance with the unified form of geometric phase of light $\Phi_G = -\hbar^{-1}\int \mathbf{S} \cdot \mathbf{\Omega}_\varphi \, d\varphi$ (ref. 11), and here $\mathbf{S} = 2\sigma/(1+\sigma^2)\hbar \cdot \mathbf{z} = S_z \cdot \mathbf{z}$ is the SAM and $\mathbf{\Omega}_\varphi = -\mathbf{z}$ is the angular velocity of coordinate rotation with respect to coordinate $\varphi$ for CCW WGMs (see Figure 1c). For CW WGMs, $\mathbf{\Omega}_\varphi = \mathbf{z}$ and the geometric phase becomes $\Phi_G = -2\sigma\varphi/(1+\sigma^2)$ (see Supplementary Figure 1).

On the other hand, it's interesting to find that the z-compnent of TAM in CVVs ($J_z = L_z + S_z = l_{TC}\hbar$) is conserved with the given WGM azimuthal mode order $p$ and grating number $q$, regardless of the transverse-spin state. This is attributed to the rotationally symmetric 'anisotropy' orientation of the scatterer group[48], and consequently the net transfer of AM between the WGMs (carrying TAM of $p\hbar$ per photon) and device is constantly $-q\hbar$. More importantly, the transverse-spin dependent SOI can be identified in Equation (5), and by engineering the transverse-spin state $\sigma$ and consequently the transverse-spin dependent geometric phase $\Phi_G$, the OAM state of a CVV can be modulated and partially converted with SAM. This is a new type of SOI, and the first manifestation of spin-to-orbital AM conversion in optical vortices directly stemming from the transverse spin of light. In addition, the left- and right-hand CP vortices in Equation (4) possess the topological charges of $l_{TC}-1$ and $l_{TC}+1$, respectively. The composition of this 'superposition' is subject to the transverse-spin state of WGM evanescent wave. Particularly, when the polarization at the grating scatterer locations reaches one of the CP states (i.e., $\sigma = \pm 1$), this superposition reduces to a single CP scalar vortex state with a single OAM eigen-state ($l = l_{TC} \mp 1$).

It should be mentioned that, by exciting WGMs from the alternative waveguide ports or scattering the evanescent waves on the other side of resonator waveguide, the sign of the transverse spin will be flipped, and using the alternative waveguide port to excite CW propagating WGMs will also reverse the sign of $l_{TC}$ (see Supplementary Note 1). Nevertheless, the general SOI phenomena and mode decomposition described in Equations (4) and (5) still hold.

**Transverse spin engineering.** The transverse-spin state $\sigma$ in the WGM evanescent wave is dependent on the ratio of cylindrical components as shown in Equation (3). In contrast to the evanescent waves of WGMs in bottle micro-resonators[49] and unbounded evanescent waves at optical interfaces[26], where this ratio is largely determined by the refractive index contrast and the incident angle of light, the transverse spin of evanescent waves in highly confined waveguide modes is also significantly altered by the lateral confinement conditions, especially the waveguide core dimensions. By modifying the mode profile of the transverse component in the core and its spatial derivative at the waveguide boundaries, the magnitude of the longitudinal component can be engineered[50]. In other words, by tailoring the waveguide geometry and consequently the vector components of modes, $\sigma$ can be adjusted and thus enabling the engineering of transverse spin in evanescent waves.

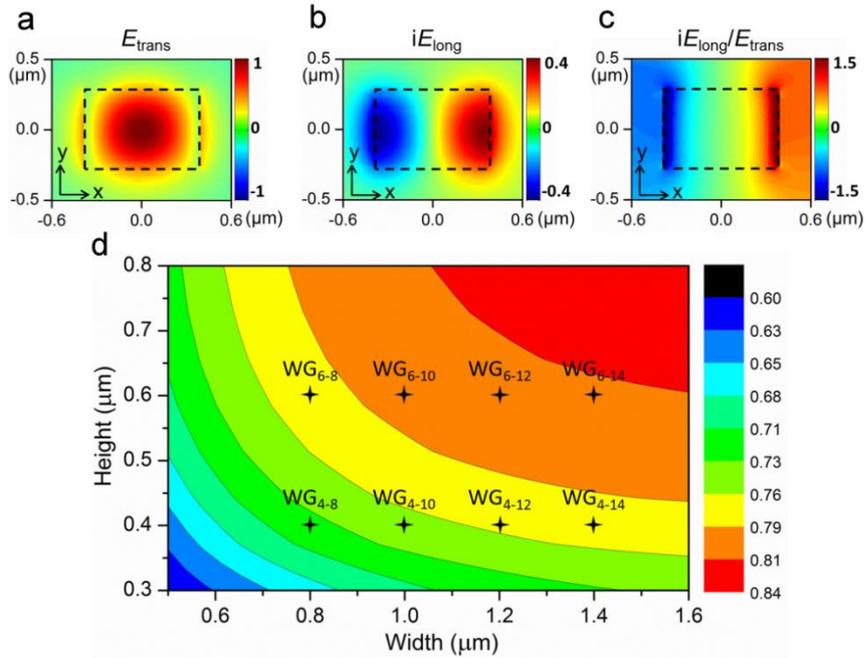

**Figure 2**. Numerically calculated field component distributions of the quasi-TE mode and the dependence of the component ratio on waveguide dimensions. (a) The cross-sectional field distribution of the transverse component $E_{trans}$ in a $SiN_x$ waveguide, and the dashed rectangular indicates the waveguide of 0.6 μm width and 0.8 μm height. The results in (b) and (c) are obtained with the same waveguide. (b) The field distribution of the longitudinal component (multiplied with the imaginary unit) $iE_{long}$. (c) The distribution of the component ratio $iE_{long}/E_{trans}$ over the waveguide cross-section and evanescent region. (d) The contour map of the ratio $iE_{long}/E_{trans}$ over variable waveguide dimensions. Among all the waveguide designs calculated, 8 waveguide dimensions marked in the map are employed for device fabrication and characterization, consisting of two different heights (0.4 and 0.6 μm) and four widths (0.8, 1.0, 1.2, and 1.4 μm) as indicated in the subscripts.

As an example, the cross-sectional maps of the fundamental quasi-TE mode components in a straight silicon nitride ($SiN_x$) waveguide (surrounded by air and placed on a $SiO_2$ substrate) is depicted in Figure 2, where the dashed rectangles indicate the waveguide cores of 0.6 μm in height and 0.8 μm in width. Apart from the transverse component $E_{trans}$ (Figure 2a), a strong longitudinal component $E_{long}$ at the core-cladding interface can also be observed in $\pm\pi/2$ phase difference to $E_{trans}$, as shown in Figure 2b. The map of the ratio $iE_{long}/E_{trans}$ is also plotted in Figure 2c, and outside the waveguide sidewalls it remains almost constant in the decaying evanescent wave, as both components decay at the same rate. More importantly, a contour map of this ratio is plotted in Figure 2d, in which an effectively variable ratio of the two components can be observed over various waveguide dimensions. Variable transverse-spin state in waveguide evanescent wave can thus be achieved with routine waveguide design[51]. The 8 waveguide designs we choose for experimental investigation are marked in the map, and their parameters are listed in Table 1. $SiN_x$ waveguide is employed for its moderate refractive index (~ 2.01) so that a larger range of transverse-spin state can be accessed than other materials (e.g., silicon).

Table 1. Design parameters of the fabricated devices

| Sample | $WG_{4-8}$ | $WG_{4-10}$ | $WG_{4-12}$ | $WG_{4-14}$ | $WG_{6-8}$ | $WG_{6-10}$ | $WG_{6-12}$ | $WG_{6-14}$ |
|---|---|---|---|---|---|---|---|---|
| Waveguide Height ($\mu$m) | 0.4 | 0.4 | 0.4 | 0.4 | 0.6 | 0.6 | 0.6 | 0.6 |
| Waveguide Width ($\mu$m) | 0.8 | 1.0 | 1.2 | 1.4 | 0.8 | 1.0 | 1.2 | 1.4 |

*These parameters apply both to the ring waveguide and access waveguide.

*The ring radius of all sample devices is 80 μm, gap between ring and access waveguide is 200 nm, and each square-shape scatterer is 100 nm by 100nm (with the same height as waveguide).

For the 8 sample devices, the ring radius of 80 μm is used. For each device, $q = 517$ scatterers are embedded on the inner-sidewall of ring. The ratio of evanescent cylindrical components may be perturbed by the presence of scatterers in the evanescent region, as represented by matrix $\mathbf{M}_1$. In this proof-of-principle study, we consider square-shape scatterers protruding from the waveguide sidewall. Each scatterer has the constant area of 100 nm by 100 nm, but is in the same height as the ring waveguide. The gap between the access waveguide and ring resonator is fixed at 200 nm. The calculated square of transverse-spin state ($\sigma^2$) of all sample devices over the scatterer region is shown in Figure 3a. A wide $\sigma^2$ range of 0.41 - 0.97 is predicted. $\sigma > 0$ holds for all cases with WGMs excited by injecting light into Port 1 and the evanescent wave at the inner sidewall is left-hand elliptical-polarized. Especially, near-circular transverse spin is expected from the devices $WG_{6-8}$ and $WG_{4-10}$ with $\sigma^2 \approx 0.95$ and 0.97, respectively. Some scanning electron microscope (SEM) images of device $WG_{6-8}$ are shown in Figure 3b and 3c.

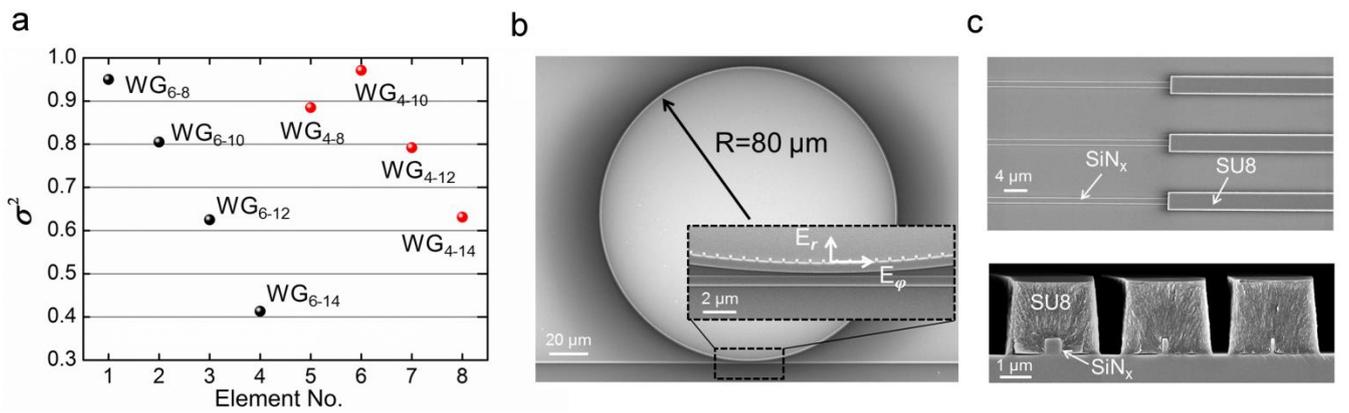

**Figure 3**. Calculated transverse-spin states of all designed devices and SEM images of fabricated device $WG_{6-8}$. (a) Calculated squared transverse-spin states in the evanescent wave of all 8 designed devices. These results are obtained considering that the WGM is excited by injection from Port 1. (b) SEM image of the device $WG_{6-8}$. The inset shows a close-up of the coupling section between the access waveguide and the resonator. (c) Top: junction point of the tapered coupler consisting of a tapered $SiN_x$ waveguide and a SU8 waveguide. Bottom: cross-section views at various positions of the tapered coupler. The minimum width of the $SiN_x$ taper (shown in the right-hand side image) is 130 nm.

**Polarization and transverse-spin state characterization.** Firstly, the average 'cylindrical' polarization ellipticity of the CVVs is measured to show the overall effect of near-field transverse spin on the polarization of far-field CVVs. The polarization of CVVs varies in space but exhibits a cylindrical symmetry with respect to the propagation axis[43], and therefore here the components $E_r$ and $E_\varphi$ are measured to characterize the average ellipticity in the cylindrical basis (i.e., $\varepsilon = |E_r|/|E_\varphi|$ or $|E_\varphi|/|E_r|$), and compared with the calculated near-field transverse-spin state which is also defined in the same basis. A Radial Polarization Convertor (RPC) is used to convert $E_r$ and $E_\varphi$ in far-field CVVs into x- and y-polarized fields respectively[52], and the power of these two components ($P_r$ and $P_\varphi$) is then recorded for $\varepsilon^2$ calculation ($\varepsilon^2 = P_r/P_\varphi$ or $P_\varphi/P_r$) (see Supplementary Note 4).

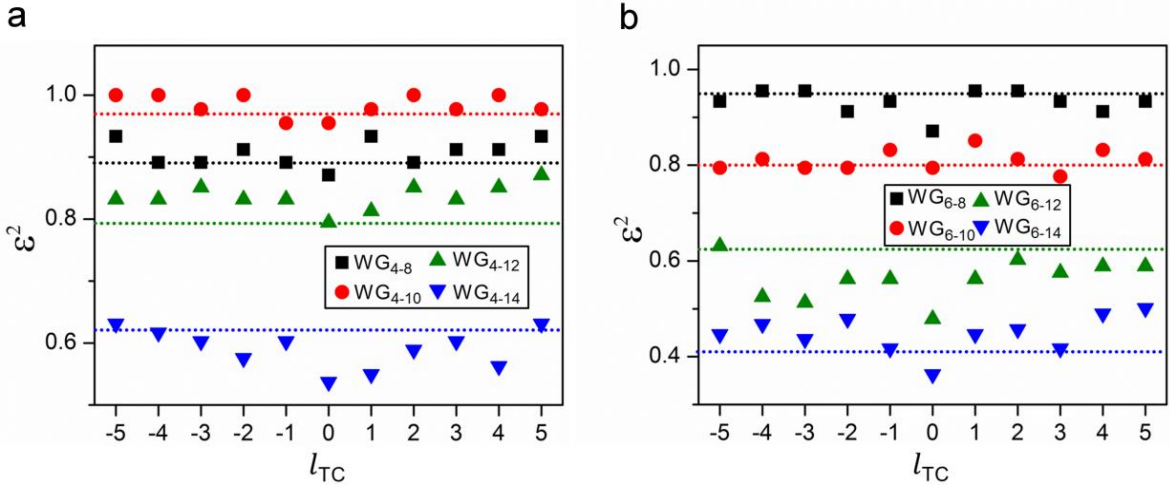

**Figure 4**. Characterization of average polarization state in CVVs. (a, b) Measured squared polarization ellipticity $\varepsilon^2$ (solid markers) of the CVVs from the devices of height 0.4 μm and 0.6 μm, respectively. The prediction from numerical calculations (plotted in Figure 3a) is plotted with dashed lines, and the measured and calculated results for the same device are marked in the same colour.

The measured $\varepsilon^2$ in CVVs of various $l_{TC}$ from all devices is shown in Figure 4 as solid markers, while the corresponding predicted $\sigma^2$ of each device from Figure 3a is plotted as the dashed line in the same color. Overall, the measured $\varepsilon^2$ exhibits high uniformity over all $l_{TC}$. CVVs of a wide range of spin states ($\varepsilon^2$ from ~0.4 to ~1.0) is obtained with various waveguide designs, and the agreement between the measured $\varepsilon^2$ and calculated $\sigma^2$ shows a definitive correspondence from the transverse-spin state in guided evanescent waves to the polarization in emitted vortices ($\varepsilon^2 = \sigma^2$). Particularly, near-CP ($|\sigma| \approx 1$) CVVs are observed with devices WG$_{4\text{-}10}$ and WG$_{6\text{-}8}$, indicating that the reduced superposition of single spin-orbital eigen-state vortices predicted by Equation (4) can be reached.

Secondly, Stokes polarimetry is performed to characterize the local transverse-spin state distribution in near-field CVVs. With the Jones vector shown in Equation (4), the normalized Stokes parameters as a function of the azimuthal coordinate can be obtained as[53]

$$S_1 = \frac{1-\sigma^2}{1+\sigma^2}\cos 2\varphi, \quad S_2 = \frac{1-\sigma^2}{1+\sigma^2}\sin 2\varphi, \quad S_3 = \frac{2\sigma}{1+\sigma^2} \quad (7)$$

For a device of a larger $|\sigma|$, the trajectory of the Stokes vector [$S_1$, $S_2$, $S_3$] on the Poincare sphere circles the pole twice at a higher latitude parallel to the equator. For $|\sigma| = 1$, the circle contracts to a single point at the poles, producing a CP CVV.

The measured Stokes parameters of near-field CVVs are depicted in Figure 5, in which the results of $l_{TC} = +4$ CVVs from the devices WG$_{6\text{-}8}$, WG$_{6\text{-}10}$, WG$_{6\text{-}12}$, and WG$_{6\text{-}14}$ are shown, respectively (see Supplementary Note 4 for details). In each case, panel (i) is the measured near-field intensity. For better comparison, the calculated Stokes parameters from Equation (7) are shown as solid curves in panel (v) in each case, along with the measured values (dots) sampled from the corresponding parameter panels of (ii)-(iv). The $\sigma$ values used for calculations are imported from Figure 3a, and the measured $S_1$, $S_2$, and $S_3$ are sampled from the pixels on the periphery of the near-field circle of 80 μm radius (i.e., the radius of ring resonator). The agreement between the theoretical curves and measured dots shown in the $S_3$ plots of (v) validates the overall effect of waveguide geometry on the transverse-spin state in evanescent waves. For devices of larger $|\sigma|$, e.g., WG$_{6-8}$, $S_1$ and $S_2$ oscillate less, indicating local polarization states of larger ellipticity.

Generally, the jitters in the measured results are attributed to the non-uniformity of fabricated gratings, as well as the decaying intensity of WGMs along the resonator. The deviation of measurements from the theory is more evident with devices of smaller $|S_3|$. This is possibly caused by the light that is scattered from the other (outer-) side of waveguide, carrying the opposite $\sigma$, due to sidewall roughness. In some devices, standing-wave-like patterns (e.g., map (iv) in Figure 5c) are introduced by the interference of scattered TE and TM modes, because in these waveguide designs these two modes are more degenerate and single-polarization-mode excitation is more critical to polarization control in mode launching.

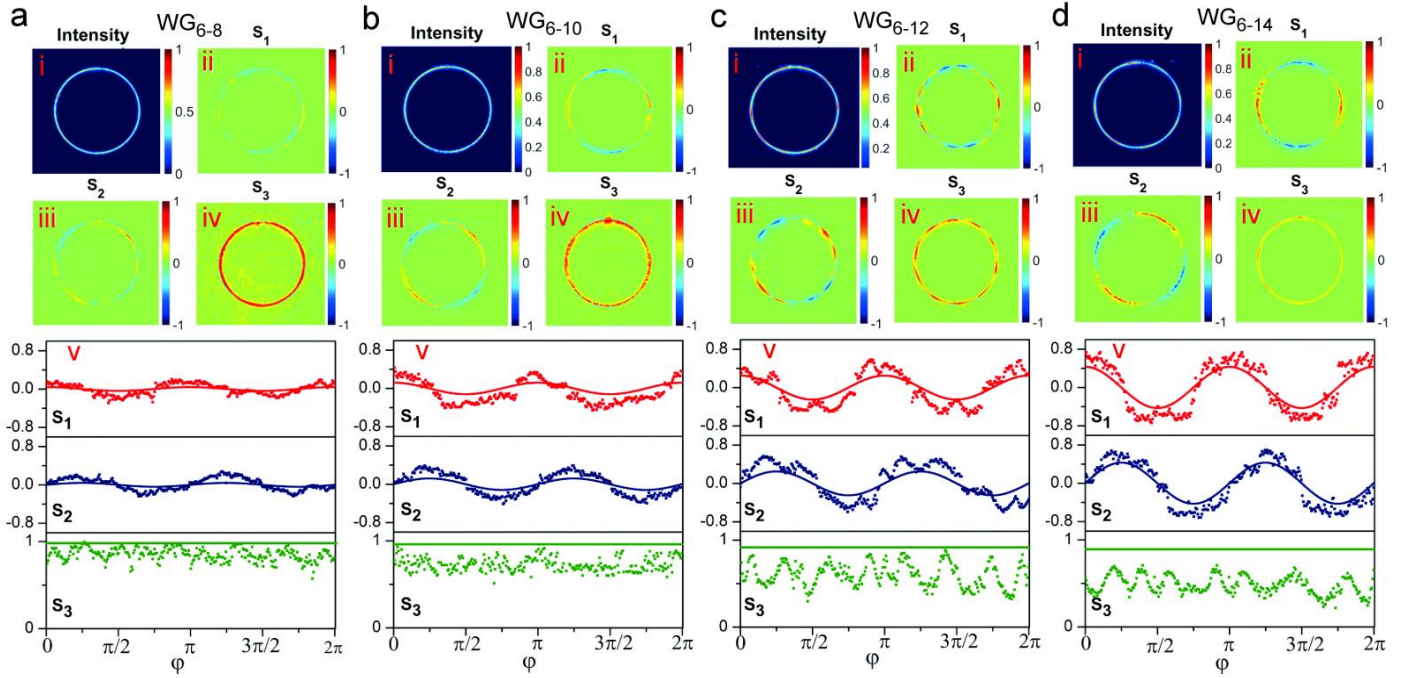

**Figure 5**. Stokes polarimetry of near-field polarization of CVVs. (a-d) Measured two-dimensional maps of near-field Stokes parameters and the comparison with theoretical prediction for devices WG$_{6-8}$, WG$_{6-10}$, WG$_{6-12}$, and WG$_{6-14}$, respectively. Each map (i) is the near-field intensity profile from the device with $l_{TC} = +4$. Maps (ii), (iii), and (iv) are the corresponding near-field profiles of the normalized Stokes parameters $S_1$, $S_2$, and $S_3$, respectively. The plots in (v) show the comparison between the measured results (dots) sampled from (ii)-(iv) and the corresponding prediction (solid curves) from Equation (7). For each set of measured data in (v), 288 pixels intersecting with the circle of 80 μm radius along the azimuthal direction ($\varphi$) from 0 to $2\pi$ are sampled from the corresponding map. For each solid curve of prediction, the data is calculated by substituting the transverse-spin state $\sigma$ from Figure 3a into Equation (7).

**Transverse spin induced SOI.** The OAM component carried by CVVs is measured to verify the transverse-spin induced spin-to-orbital conversion predicted by Equation (5) (see Supplementary Note 4 and 5 for characterization method of OAM state and emission spectrum from devices). The measured OAM spectra for the CVVs from the devices WG$_{6-8}$, WG$_{6-10}$, WG$_{6-12}$, and WG$_{6-14}$ are plotted in Figure 6. In close agreement with the theory, each OAM spectrum (row) of CVV with $l_{TC}$ contains two dominant peaks at $l_{TC}-1$ and $l_{TC}+1$, carried by the constituent left- and right-hand CP vortices, respectively. The intensities of all spurious modes are < 0.03. Note that each CP vortex can thus be confirmed as possessing a TAM of $l_{TC}\hbar$ (see Supplementary Note 6), and this experimentally validates the overall TAM in each CVV is preserved as $l_{TC}\hbar$ regardless of waveguide geometries. More importantly, the average SAM in each CVV is subject to the near-field transverse spin ($S_z = 2\sigma/(1+\sigma^2)\hbar$) as shown in Figure 5. Therefore, the remarkable transverse-spin dependent SOI effect is revealed, as the OAM component carried by CCVs can be partially derived out of the transverse SAM in the evanescent waves. This is the first demonstration of an SOI effect resulted from the interaction between the intrinsic OAM and transverse SAM of light.

A direct and useful manifestation of this effect is that the relative intensities of the two dominant peaks, i.e., the two constituent CP vortices, can be changed by modifying $\sigma$. For example, the normalized intensities of the left- and right-hand CP vortices from WG$_{6-8}$ are around 0.93 and 0.07, respectively, while for WG$_{6-14}$ they account for about 0.62 and 0.36 of the total intensity, respectively. This variable superposition of AM states in CVVs provides a viable pathway for information encoding in the spin-orbit space. Another implication of this SOI effect is that a vortex should appear even when $l_{TC} = 0$ but $\sigma \neq 0$ (exemplified by the square in the yellow box in Figure 6a); that is, without introducing any spatial phase gradient that has been inherent to many optical vortex generation techniques[5]. This purely transverse-spin-derived vortex essentially originates from the spatially varying 'anisotropy' of the gratings and the rotational symmetry of vector WGMs. In other words, this is an interesting demonstration of optical vortex generation controlled by the QSHE of light[33], and the spin state in the edge modes stemming from the intrinsic SOI at optical interfaces can thus be manipulated for spatial light modulation via the 'extrinsic' SOI in anisotropic structures[11].

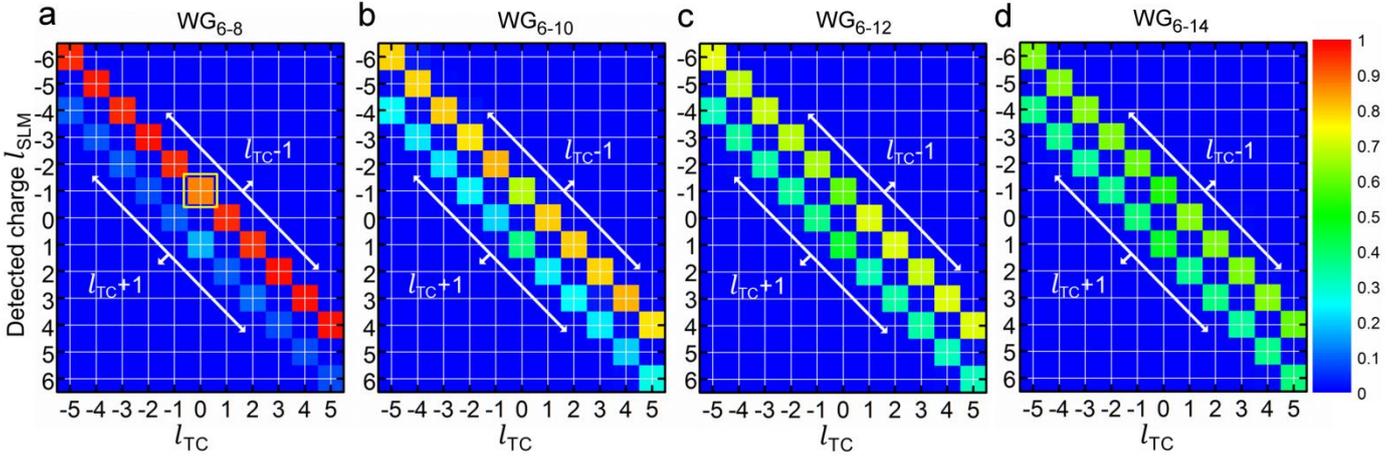

**Figure 6**. Characterization of OAM components in CVVs. (a-d) The measured OAM spectra for the devices WG$_{6-8}$, WG$_{6-10}$, WG$_{6-12}$, and WG$_{6-14}$, respectively. For each device, the wavelengths of $l_{TC}$ = -5 to +5 are considered, and each column represents a spectrum of measured OAM components with the corresponding $l_{TC}$.

**Spin-orbit controlled unidirectional coupling.** Given the principle of reciprocity, this device can also be used for detection of AM components in an incident CVV beam[54]. The ring resonator supports the degenerate CW and CCW WGMs at each resonance wavelength $\lambda_0$, and these two modes give rise to the emission of two CVVs of opposite TCs, i.e., $l_{TC} = \pm(p - q)$. Meanwhile, these two WGMs exhibit opposite $\sigma$ in the inner-side evanescent waves, and therefore the two emitted CVVs carry exactly opposite spin and orbital AM states, i.e., <$2\sigma/(1+\sigma^2)$, $l_{TC} - 2\sigma/(1+\sigma^2)$> and <$-2\sigma/(1+\sigma^2)$, $l_{TC} + 2\sigma/(1+\sigma^2)$>. When receiving at $\lambda_0$, this device can couple these two CVVs into the two opposite resonating directions and guide their power to the two access-waveguide ports, respectively. All CVVs with $\lambda \neq \lambda_0$, or at $\lambda_0$ but with other SAM and OAM states are mismatched with this selection rule and will be denied by the device. Consequently, we obtain the effect of unidirectional coupling into guided modes jointly controlled by spin and orbital AM states. Although this phenomenon is essentially associated with the spin-direction locking induced by local SOI in evanescent waves of guided modes[21,22], this new spin-orbit direction locking effect incorporates the orbital degree of freedom, using the close-loop waveguide for filtering in the OAM space. This spin-orbit controlled coupling provides a potential solution for spin and orbital AM state detection, avoiding the separate manipulations on these two degrees of freedom.

Generally, the input light for ideal reception with the device should carry the identical spin and orbital AM states as the output CVVs, while exhibiting cylindrical symmetry in intensity and polarization profiles. But for this proof-of-principle study here, the special case of $\sigma = \pm 1$ (where the spin-orbit states of CVVs reduce to $<\pm\sigma, l_{TC}\mp\sigma>$) is demonstrated for simpler experimental configuration, using the device $WG_{4-10}$ of near-CP transverse-spin state ($\sigma \approx \pm 1$) as shown in Figure 4b.

Table 2. SAM and OAM states in CVVs at various resonance wavelengths

| Wavelength (nm) | 1578.61 | | 1583.11 | | 1587.59 | | 1592.11 | |
|---|---|---|---|---|---|---|---|---|
| Access port | 1 | 2 | 1 | 2 | 1 | 2 | 1 | 2 |
| $l_{TC}$ | −4 | +4 | −2 | +2 | 0 | 0 | +2 | −2 |
| SAM | −1 | +1 | −1 | +1 | −1 | +1 | −1 | +1 |
| OAM | −3 | +3 | −1 | +1 | +1 | −1 | +3 | −3 |

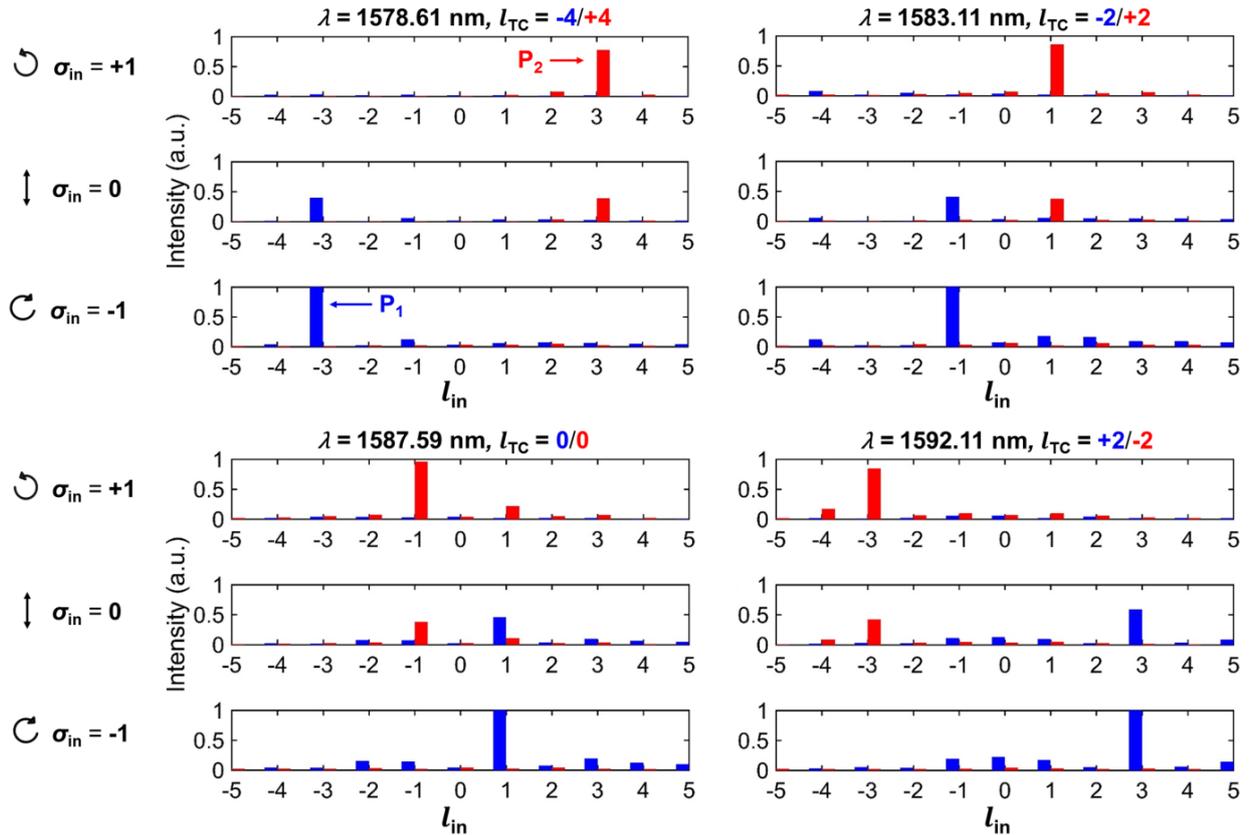

**Figure 7**. Proof-of-principle illustration of spin-orbit controlled uni-directional coupling of waveguide modes. (a-d) Measured results for the device $WG_{4-10}$ with incident light in the wavelength of 1578.61 nm, 1583.11 nm, 1587.59 nm, and 1592.11 nm, respectively. For each wavelength, incident beams of 33 different spin-orbit states ($\sigma_{in} = -1$, 0, and +1, $l_{in} = -5, -4, \ldots,$ and +5) are illuminated on the device. For each incident polarization state ($\sigma_{in}$ = −1/0/+1), the received power with different incident OAM orders from the both ports are listed in a single histogram. The data in each figure (a/b/c/d) is normalized to the highest value in the group.

The spin and orbital AM states of the CVVs at four resonant wavelengths from device WG$_{4\text{-}10}$ associated to Port 1 or 2 are listed in Table 2. For each wavelength, optical vortices of 3 SAM states ($\sigma_{in}$ = 0, ±1) and 11 OAM states ($l_{in}$ from −5 to +5) are prepared and illuminated on the device (see Supplementary Note 6 for details). The measured (and calibrated with respect to the lensed fiber coupling loss) output power P$_1$ at port 1 and P$_2$ at port 2 are normalized and plotted in Figure 7. The first distinctive observation is that P$_1$ in blue bars (P$_2$ in red bars) is universally negligible with incidence of $\sigma_{in}$ = +1 (−1), in accordance with the predefined transverse-spin state $\sigma \approx$ −1 (+1) when inputting via Port 1 (2) and the underlying prediction from the spin-direction locking effect[21,22]. With the incidence of an arbitrary polarization state, however, light is coupled to the both ports and the resulting ratio of P$_1$ and P$_2$ is determined by the relative intensity of left- and right-hand CPs in the incident CVV. For example, with the incident linear polarization ($\sigma_{in}$ = 0, middle rows in Figure 7a to 7d) as an equal superposition of two CPs, P$_1$ and P$_2$ exhibit comparable values. Moreover, the coupling strength is further subject to the incident OAM state $l_{in}$. For example, when measuring at Port 2 with incidence at 1578.61 nm, a single dominant power peak at port 2 appears only at the incident state of < $\sigma_{in}$ = +1, $l_{in}$ = +3> (upper row in Figure 7a), while it can only be observed at Port 1 with incident <$\sigma_{in}$ = −1, $l_{in}$ = −3> (lower row in Figure 7a). This highly directional and selective coupling, determined by the spin and orbital AM state <$\sigma_{in}$, $l_{in}$>, is a higher-order phenomenon with respect to the basic spin-controlled coupling via evanescent waves, as both the spatial and polarization properties of light must be taken into account. This effect allows for a robust manipulation of light on the micron-scale using both the spin and orbital degrees of freedom, e.g., encoding and retrieving information, without the necessity of separate controls on polarization and spatial phase profile.

## Discussion

To sum up, we have identified and demonstrated a new type of spin-orbit interaction of light, namely the interplay between the intrinsic OAM and the transverse spin of light. This new SOI effect originates from the manipulation of local transverse-spin-dependent geometric phase by artificially introducing a close-loop waveguide and sub-wavelength scatterers of rotational symmetry. Engineering the local transverse spin by tailoring waveguide dimensions then controls the global spin-to-orbital conversion in the generated optical vortices.

Our results have both fundamental and applied importance. The interaction between the intrinsic OAM and transverse spin of light is an integral but thus far missing part of the rich SOI phenomena. The newly discovered interaction builds one more pathway between the polarization and spatial degrees of freedom of light, which could provide nano-photonic technologies with additional tools of light manipulation at the subwavelength scales and of information transfer over more degrees of freedom. The resulting effects, e.g., the variable superposition of spin-orbit states in optical vortices, may find applications in optical quantum information processing. The spin-orbit jointly controlled directional coupling can be used to operate on the eigen-states involving both AM components, so that the device considered here can be regarded as a prototype of a planar spin-orbit-controlled gate that interfaces propagating and bounded photons of two-dimensional entanglement. Better performance (e.g., power efficiency and AM state purity) can be brought about by further device design and optimization. The demonstrated interaction should also exist in other systems that support evanescent modes, including surface plasmon-polaritons which can significantly miniaturize the elements.

## Methods

**Numerical simulations.** Numerical simulations are performed with the finite difference eigenmode solver (FDE, Lumerical Solutions, Inc.). For the calculation of squared transverse-spin state ($\sigma^2$ shown in Figure. 3a) at the scatterer location of each designed device, first the distribution of the two cylindrical components ($E_r$ and $E_\varphi$) over the scatterer region is calculated, and then the average $\sigma^2$ (<=1) in the scattered evanescent wave is obtained as the ratio of integrated intensities of $E_r$ and $E_\varphi$ over one scatterer region. The effect of sactterer's modulation on field amplitudes ($W_1/W_2$ in Equation (3)) is thus included in the calculated $\sigma^2$.

**Fabrication.** The $SiN_x$ waveguide layers are first deposited on a 5-μm oxidized <100> silicon wafer using inductively coupled plasma chemical vapor deposition (ICP-CVD) system (Plasmalab System 100 ICP180, Oxford). The device structures are defined in a 450-nm-thick negative resist using electron-beam lithography (EBL, EBPG5000 ES, Vistec). Reactive-ion-etch (RIE, Plasmalab System 100 RIE180, Oxford) with a mixture of $CHF_3$ and $O_2$ gases is applied to etch through the waveguide layer to form the device. An inverse taper combined with a SU8 waveguide is used as the coupler between external optical fiber and the access waveguide.

**Experimental setups** for device characterizations, SOI measurement and spin-orbit controlled unidirectional coupling are shown and explained in Supplementary Note 4 and 6.

## References

1. Poynting, J. H. The wave motion of a revolving shaft, and a suggestion as to the angular momentum in a beam of circularly polarised light. *Proc. R. Soc. Lond. Ser. A* **82**, 560–567 (1909).
2. Beth, R. A. Mechanical detection and measurement of the angular momentum of light. *Phys. Rev.* **50**, 115-125 (1936).
3. Allen, L., Beijersbergen, M. W., Spreeuw, R. J. C. & Woerdman, J. P. Orbital angular momentum of light and the transformation of Laguerre-Gaussian laser modes. *Phys. Rev. A* **45**, 8185-8189 (1992).
4. Allen, L., Padgett, M. J. & Babiker, M. The orbital angular momentum of light. *Prog. Opt.* **39**, 291-372 (1999).
5. Yao, A. M. & Padgett, M. J. Orbital angular momentum: origins, behavior and applications. *Adv. Opt. Photon* **3**, 161-204 (2011).
6. Bliokh, K. Y. & Nori, F. Transverse and longitudinal angular momenta of light. *Phys. Rep.* **592**, 1-38 (2015).
7. Andrews, D. L., & Babiker, M. *The angular momentum of light* (Cambridge Univ. Press, Cambridge, 2013).
8. Liberman, V. S. & Zel'dovich, B. Y. Spin-orbit interaction of a photon in an inhomogeneous medium. *Phys. Rev. A* **46**, 5199-5207 (1992).
9. Hasman, E., Biener, G., Niv, A. & Kleiner, V. Space-variant polarization manipulation. *Prog. Opt.* **47**, 215-289 (2005).
10. Marrucci, L. *et al*. Spin-to-orbital conversion of the angular momentum of light and its classical and quantum applications. *J. Opt.* **13**, 064001 (2011).
11. Bliokh, K. Y., Rodriguez-Fortuno, F. J., Nori, F. & Zayats, A. V. Spin-orbit interactions of light. *Nat. Photon* **9**, 796-808 (2015).
12. Schwartz, C. & Dogariu, A. Conservation of angular momentum of light in single scattering. *Opt. Express* **14**, 8425-8433 (2006).
13. Zhao, Y., Edgar, J. S., Jeffries, G. D. M., McGloin, D. & Chiu, D. T. Spin-to-orbital angular momentum conversion in a strongly focused optical beam. *Phys. Rev. Lett*. **99**, 073901 (2007).
14. Ciattoni, A., Cincotti, G. & Palma, C. Angular momentum dynamics of a paraxial beam in a uniaxial crystal. *Phys. Rev. E* **67**, 036618 (2003).
15. Bliokh, K. Y., Niv, A., Kleiner, V. & Hasman, E. Geometrodynamics of spinning light. *Nat. Photon* **2**, 748-753 (2008).
16. Bliokh, K. Y. Geometrical optics of beams with vortices: Berry phase and orbital angular momentum Hall effect. *Phys. Rev. Lett*. **97**, 043901 (2006).
17. Hosten, O. & Kwiat, P. Observation of the spin Hall effect of light via weak measurements. *Science* **319**, 787-790 (2008).
18. Rodriguez-Fortuno, F. J. et al. Near-Field Interference for the Unidirectional Excitation of Electromagnetic Guided Modes. *Science* **340**, 328-330 (2013).
19. Lin, J. *et al*. Polarization-Controlled Tunable Directional Coupling of Surface Plasmon Polaritons. *Science* **340**, 331-334 (2013).
20. Shitrit, N. *et al*. Spin-Optical Metamaterial Route to Spin- Controlled Photonics. *Science* **340**, 724-726 (2013).
21. O'Connor, D., Ginzburg, P., Rodriguez-Fortuno, F. J., Wurtz, G. A. & Zayats, A. V. Spin-orbit coupling in surface plasmon scattering by nanostructures. *Nat. Commun.* **5**, 5327 (2014).
22. Petersen, J., Volz, J. & Rauschenbeutel, A. Chiral nanophotonic waveguide interface based on spin-orbit interaction of light. *Science* **346**, 67-71 (2014).
23. Roy, B., Ghosh, N., Banerjee, A., Gupta, S. D. & Roy, S. Manifestations of geometric phase and enhanced spin Hall shifts in an optical trap. *New J. Phys.* **16**, 083037 (2014).

24. Rodriguez-Herrera, O. G., Lara, D., Bliokh, K. Y., Ostrovskaya, E. A. & Dainty, C. Optical Nanoprobing via Spin-Orbit Interaction of Light. *Phys. Rev. Lett.* **104**, 253601 (2010).
25. Yu, N. & Capasso, F. Flat optics with designer metasurfaces. *Nat. Mater.* **13**, 139-150 (2014).
26. Bliokh, K. Y. & Nori, F. Transverse spin of a surface polariton. *Phys. Rev. A* **85**, 061801 (2012).
27. Kim, K.-Y., Lee, I.-M., Kim, J., Jung, J. & Lee, B. Time reversal and the spin angular momentum of transverse-electric and transverse-magnetic surface modes. *Phys. Rev. A* **86**, 063805 (2012).
28. Bliokh, K. Y., Bekshaev, A. Y. & Nori, F. Extraordinary momentum and spin in evanescent waves. *Nat. Commun.* **5**, 3300 (2014).
29. Kim, K.-Y. & Wang, A. X. Spin angular momentum of surface modes from the perspective of optical power flow. *Opt. Lett.* **40**, 2929-2932 (2015).
30. Aiello, A., Banzer, P., Neugebaueru, M. & Leuchs, G. From transverse angular momentum to photonic wheels. *Nat. Photon.* **9**, 789-795 (2015).
31. Banzer, P. *et al.* The photonic wheel - demonstration of a state of light with purely transverse angular momentum. *J. Europ. Opt. Soc. Rap. Public* **8**, 13032 (2013).
32. Born, M. & Wolf, E. Principles of optics Ch. 6 (Pergamon Press, Oxford, 1980).
33. Bliokh, K. Y., Smirnova, D. & Nori, F. Quantum spin Hall effect of light. *Science* **348**, 1448-1451 (2015).
34. Ling, X. *et al.* Giant photonic spin Hall effect in momentum space in a structured metamaterial with spatially varying birefringence. *Light Sci. Appl.* **4**, e290 (2015).
35. Ling, X. *et al.* Recent advances in the spin Hall effect of light. *Rep. Prog. Phys.* **80**, 066401 (2017).
36. Young, A. B. *et al.* Polarization Engineering in Photonic Crystal Waveguides for Spin-Photon Entanglers. *Phys. Rev. Lett.* **115**, 153901 (2015).
37. Sayrin, C. *et al.* Optical diode based on the chirality of guided photons. arXiv, 1502.01549 (2015).
38. Pichler, H., Ramos, T., Daley, A. J. & Zoller, P. Quantum optics of chiral spin networks. *Phys. Rev. A* **91**, 042116 (2015).
39. Lodahl, P. *et al.* Chiral quantum optics. *Nature* **541**, 473-480 (2017).
40. Rubinsztein-Dunlop, H. *et al.* Roadmap on structured light. *J. Opt.* **19**, 013001 (2017).
41. Cai, X. *et al.* Integrated Compact Optical Vortex Beam Emitters. *Science* **338**, 363-366 (2012).
42. Zhu, J., Chen, Y., Zhang, Y., Cai, X. & Yu, S. Spin and orbital angular momentum and their conversion in cylindrical vector vortices. *Opt. Lett.* **39**: 4435-4438 (2014).
43. Zhu, J., Cai, X., Chen, Y. & Yu, S. Theoretical model for angular grating-based integrated optical vortex beam emitters. *Opt. Lett.* **38**, 1343-1345 (2013).
44. Bomzon, Z., Biener, G., Kleiner, V. & Hasman, E. Space-variant Pancharatnam-Berry phase optical elements with computer-generated subwavelength gratings. *Opt. Lett.* **27**, 1141-1143 (2002).
45. Neugebauer, M., Bauer, T., Aiello, A. & Banzer, P. Measuring the Transverse Spin Density of Light. *Phys. Rev. Lett.* **114**, 063901 (2015).
46. Marcuvitz, N. *Waveguide Handbook 1st edn* (McGraw-Hill, New York, 1951).
47. Pancharatnam, S. Proc. Indian Acad. Sci. Sect. A **44**, 247 (1956).
48. Marrucci, L., Manzo, C. & Paparo, D. Optical spin-to-orbital angular momentum conversion in inhomogeneous anisotropic media. *Phys. Rev. Lett.* **96**, 163905 (2006).
49. Junge, C., O'Shea, D., Volz, J. & Rauschenbeutel, A. Strong Coupling between Single Atoms and Nontransversal Photons. *Phys. Rev. Lett.* **110**, 213604 (2013).
50. Driscoll, J. B. *et al.* Large longitudinal electric fields (Ez) in silicon nanowire waveguides. *Opt. Express* **17**, 2797-2804 (2009).
51. Snyder, A. W. & Love, J. *Optical waveguide theory*. (Springer Science & Business Media, Berlin, 2012).
52. Stalder, M. & Schadt, M. Linearly polarized light with axial symmetry generated by liquid-crystal polarization converters. *Opt. Lett.* **21**, 1948-1950 (1996).
53. Berry, H. G., Gabrielse, G. & Livingston, A. E. Measurement of the Stokes parameters of light. *Appl. Opt.* **16**, 3200-3205 (1977).
54. Cicek, K. *et al.* Integrated optical vortex beam receivers. *Opt. Express* **24**, 28529-28539 (2016).

**Supplementary Note 1. Formulation of Cylindrical Vector Vortices Emission**

The evanescent wave of whispering-gallery modes (WGMs), written in the cylindrical polarization basis as $\mathbf{E}_{in} \propto e^{\pm ip\varphi}[E_r \ E_\varphi]^T$, is perturbed by the second-order grating when circulating around the resonator. Here we denote the positive integer $p$ as the azimuthal mode number of WGMs, and the two degenerate counter-propagating WGMs resonating in the same wavelength have the mode numbers of $p$ (counter-clockwise, CCW) and $p$ (clockwise, CW), respectively. The perturbation of gratings to the evanescent wave is generalized in a matrix as

$$\mathbf{M}_1 = \begin{bmatrix} W_1 & 0 \\ 0 & W_2 \end{bmatrix} \cdot e^{i\delta(\varphi)} \quad (1)$$

where $W_1$ and $W_2$ are real numbers that reflect the modulation on the amplitudes of local transverse ($E_r$) and longitudinal ($E_\varphi$) fields, respectively, due to grating perturbation. The off-diagonal elements of $\mathbf{M}_1$ are vanishing as we assume the scattering does not introduce coupling between orthogonal field components. $\delta(\varphi) = \mp q\varphi$ is the phase imparted on the first-order diffracted wave derived using the coupled-mode theory (cf. supplementary material of ref. [1]) and $q$ the number of grating elements. In addition, as WGMs travel around the resonator, the vector evanescent wave experiences a rotation of local coordinates ($[E_r \ E_\varphi]^T$) with respect to the global laboratory frame ($[E_x \ E_y]^T$), as shown in Supplementary Figure 1. The effect of this rotation on the emitted CVVs (represented in the basis of $[E_x \ E_y]^T$) can be written with a single matrix $\mathbf{M}_2$ as

$$\begin{bmatrix} E_x \\ E_y \end{bmatrix} = \mathbf{M}_2 \cdot \begin{bmatrix} E_r \\ E_\varphi \end{bmatrix} = \begin{bmatrix} \cos\varphi & \mp\sin\varphi \\ \pm\sin\varphi & \cos\varphi \end{bmatrix} \begin{bmatrix} E_r \\ E_\varphi \end{bmatrix} \quad (2)$$

The final output CVV ($\mathbf{E}_{out} = \mathbf{M}_2 \mathbf{M}_1 \mathbf{E}_{in}$) can be obtained as

$$\mathbf{E}_{out} \propto \frac{1+\sigma}{2\sqrt{1+\sigma^2}} e^{i(l_{TC}-1)\varphi} \begin{bmatrix} 1 \\ i \end{bmatrix} \mp \frac{1-\sigma}{2\sqrt{1+\sigma^2}} e^{i(l_{TC}+1)\varphi} \begin{bmatrix} 1 \\ -i \end{bmatrix} \quad (3)$$

where $\sigma$ is the transverse-spin state defined in Equation (2) in the main text, and $l_{TC} = \pm(p - q)$ the topological charge. It should be emphasized here that the constituent left- ($[1 \ i]^T$) and right-hand ($[1 \ \ \ i]^T$) circular polarized vortices are out-of- and in-phase, respectively, only when following the two definitions of $\sigma$.

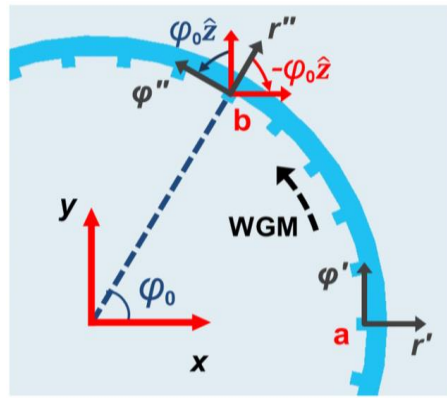

**Supplementary Figure 1.** Rotation of local coordinates as WGMs circulating around the resonator. For the counter-clockwise (CCW) propagating WGM shown here, the rotation angle of local coordinates ($r$, $\varphi$) from point **a** to **b** with respect to the global coordinates ($x$, $y$) of output CVVs is $\varphi_0\mathbf{z}$, and $\mathbf{z}$ is the unit vector. In other words, from the perspective of light field at point **b**, a coordinate rotation of angle $\varphi_0\mathbf{z}$ should be applied in order to align with the global reference frame ($x$, $y$) when its spatial (Pancharatnam) phase is compared with point **a** in the CVV field. And therefore, the angular velocity of reference frame rotation with respect to coordinate $\varphi$ for CCW WGM is $\mathbf{\Omega}_\varphi = $ z. Similarly, $\mathbf{\Omega}_\varphi = $ z for clockwise propagating (CW) WGMs.

**Supplementary Note 2. Angular Momentum in Cylindrical Vector Vortices**

The cylindrical vector vortices (CVVs) emitted from the angular-grating based devices considered in this paper exhibit good paraxiality, as the radius of ring resonator ($R = 80$ μm) is much larger than the wavelength ($\lambda = 1.5$ um) [2, 3]. The angular momentum (AM) carried in paraxial optical vortex beams can be essentially considered as the sum of the spin and orbital AM components, which are associated with the polarization and spatial properties of light, respectively [4, 5]. The cycle averaged $z$-component of the spin AM (SAM) and orbital AM (OAM) per unit length per photon of a vortex beam can be written as [5]

$$S_z = \frac{\hbar \iint r dr d\varphi \left( E_x^* E_y - E_y^* E_x \right)}{i \iint r dr d\varphi \mathbf{E}^* \cdot \mathbf{E}} \quad (4)$$

$$L_z = \frac{\hbar \iint r dr d\varphi \sum_{j=x,y,z} E_j^* \frac{\partial}{\partial \varphi} E_j}{i \iint r dr d\varphi \mathbf{E}^* \cdot \mathbf{E}} \quad (5)$$

By substituting the CVV shown in Equation (3) into the equations above, the SAM and OAM components carried by the CVV are

$$S_z = \frac{2\sigma}{1+\sigma^2} \hbar \quad (6)$$

$$L_z = \left( l_{TC} - \frac{2\sigma}{1+\sigma^2} \right) \hbar \quad (7)$$

where $\sigma$ is the transverse-spin state in the near-field evanescent wave. The total angular momentum (TAM) in a CVV ($J_z = S_z + L_z$) is thus simply written as

$$J_z = l_{TC} \hbar \quad (8)$$

**Supplementary Note 3. Geometric Phase Induced by Coordinate Rotation**

As the polarization state of CVVs is space-variant [2], here the Pancharatnam phase is used to define the phase difference of light fields in different positions in CVVs [6], that $\Phi_P = \arg\langle \mathbf{E}(r_1, \varphi_1), \mathbf{E}(r_2, \varphi_2) \rangle$, where $\arg\langle \mathbf{E}_1, \mathbf{E}_2 \rangle$ is the argument of the inner product of the two Jones vectors $\mathbf{E}_1$ and $\mathbf{E}_2$. Following this definition, the Pancharatnam phase of fields at two different positions ($r_1$, $\varphi_1$) and ($r_2$, $\varphi_2$) in a CVV is given by

$$\Phi_P = l_{TC} \Delta\varphi + \arg\left( \cos\Delta\varphi \pm i \frac{2\sigma}{1+\sigma^2} \sin\Delta\varphi \right) \quad (9)$$

where $\Delta\varphi = \varphi_2 - \varphi_1$, and the CVV excited by CCW (CW) WGM takes the + ( ) sign in the equation. The gradient of Pancharatnam phase along the azimuthal direction is

$$\lim_{\Delta\varphi \to 0} \frac{\Phi_P}{\Delta\varphi} = l_{TC} \pm \lim_{\Delta\varphi \to 0} \frac{\arctan\left( \frac{2\sigma}{1+\sigma^2} \tan\Delta\varphi \right)}{\Delta\varphi} = l_{TC} \pm \frac{2\sigma}{1+\sigma^2} \quad (10)$$

Clearly, the Pancharatnam phase in CVVs scales linearly with coordinate $\varphi$, and thus we can rewrite it as

$$\Phi_P = l_{TC}\varphi \pm \frac{2\sigma}{1+\sigma^2} \varphi \quad (11)$$

Considering the SAM component carried by CVVs shown in Supplementary Equation (6), the Pancharatnam phase can be generalized as

$$\Phi_P = l_{TC}\varphi - \frac{1}{\hbar} \int \mathbf{S} \cdot \mathbf{\Omega}_\varphi d\varphi \quad (12)$$

where $\mathbf{S} = S_z \mathbf{z}$ is the SAM per photon, and $\mathbf{\Omega}_\varphi$ is the angular velocity of reference frame rotation with respect to the coordinate $\varphi$ for Pancharatnam phase comparison (see Supplementary Figure 1). Here, $\mathbf{\Omega}_\varphi = \mp \mathbf{z}$ for CCW and CW WGMs, respectively.

**Supplementary Note 4. Techniques for Polarization and OAM States Characterization**

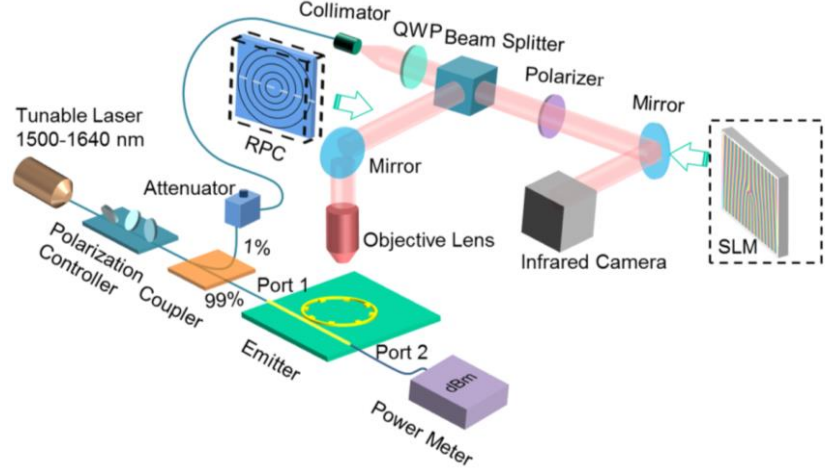

**Supplementary Figure 2**. Experimental setup for device characterization and the observation of the transverse-spin induced SOI effect.

The experimental characterizations of the devices are performed with the setup shown in Supplementary Figure 2. For the excitation of WGMs and hence emission of CVVs, the continuous-wave light from the tunable laser source (8461B, Agilent) is controlled with a fiber polarization controller (FPC561, Thorlabs), and the quasi-TE mode in the waveguide is excited by launching the horizontally polarized light into one of the ports (e.g., Port 1 as shown in Figure 3a) using a lensed fiber (SMF-28E+LL, Corning). A small fraction, 1%, of the input light is tapped using a coupler (PMC1550-90B-FC, Thorlabs) and directed to another collimator (F240FC-1550, Thorlabs) to serve as the reference light for the interference with the emitted CVVs.

For the measurement of the emission spectrum of the device, the vertically emitted beam from the device plane is collected and collimated with a 20X objective lens (UPlanFLN, Olympus) positioned in the working distance (1.7mm) away from the device. A power meter (PM122D, Thorlabs) is placed behind the collimating objective lens to record the dependence of emission power on the working wavelength, while the output wavelength of the tunable laser is swept from 1500 nm to 1640 nm with the step of 10 pm.

For measuring the average cylindrical-basis polarization ellipticity of CVVs, a liquid crystal based element called Radial Polarization Converter (RPC, ARCoptix S. A., Switzerland) is used to selectively measure the power of $E_\varphi$ and $E_r$ components. The RPC can be typically used for its spatially varying anisotropy to convert linearly polarized light into vector beams of azimuthal or radial polarizations [7]. Here the reversed effect of this element is employed: by injecting the light into the exit side, $E_\varphi$ and $E_r$ in the CVV will be converted into x- and y-polarized light leaving the entrance side, respectively. A linear polarizer (LPNIR100-MP2, Thorlabs) is then used to filter out one of the components, and by detecting the power of the two orthogonal components as $P_\varphi$ and $P_r$, the squared polarization ellipticity ($\varepsilon^2$) in the CVV determined by the near-field transverse spin state can be obtained as $\varepsilon^2 = P_r/P_\varphi$ or $P_\varphi/P_r$.

For Stokes parameters measurements, the near-field pattern of the CVV is imaged onto an InGaAs camera (C14041-10U, Hamamastu) with an achromatic lens ($f$ = 250 mm, AC254-250-C-ML, Thorlabs), and the linear- and circular-polarizations are obtained by adjusting the quarter-wave plate (QWP, AQWP 10M-1600, Thorlabs) and the linear polarizer (LP) mounted on continuous rotation mounts (CRM1, Thorlabs).

For the characterization of OAM states in CVVs, a phase-only reflective spatial light modulator (PLUTO SLM, HOLOEYE Photonics AG) loaded with grey-scale fork-grating patterns is used [8]. A linear polarizer is first used to acquire one of the linear-polarized components in the CVV, which generally is a mixture of two topologically charged vortices as shown in Equation (4) in the main text. The central axis of the polarized CVVs is then aligned with the center of fork-grating patterns on the SLM. For each incident CVV, the SLM is loaded with a series of fork-grating images with consecutive integer topological charges, e.g., $l_{SLM}$ = -5, -4, …, +5. The light reflected off each image is focused by an achromatic lens ($f$ = 150 mm, AC254-150-C-ML, Thorlabs) followed by the InGaAs camera, and the power of the corresponding OAM component $l_{SLM}$ is obtained by integrating the intensity of the central Gaussian-like spot [9]. The process is repeated for the other linear-polarized component, and the measured OAM spectrum of the incident CVV is then obtained by averaging the two corresponding OAM components over the two linear polarization components.

**Supplementary Note 5. Preliminary Characterization of Devices**

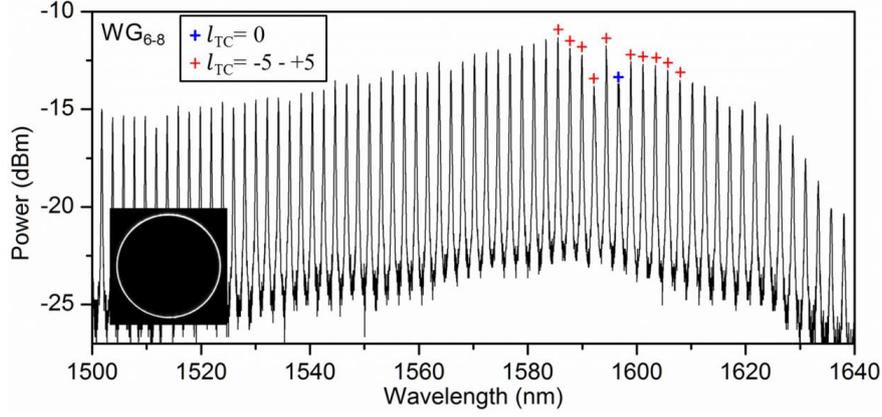

**Supplementary Figure 3.** Measured emission spectral response of sample device $W_{6-8}$ as input wavelength is swept from 1500-1640 nm. The inset shows a typical near-field intensity profile of emitted CVVs.

The measured emission spectral response of sample $WG_{6-8}$, as an instance, is plotted in Supplementary Figure 3 after normalization to the output power of tunable laser. The central wavelength at which the emitted CVV has $l_{TC} = p$ 　 $q = 0$, is $\lambda_c = 1596.6$ nm, and the free spectral range is around 2.2 nm. At the wavelengths longer (shorter) than $\lambda_c$, CVVs carry positive (negative) integer $l_{TC}$ at the resonance peaks. The inset shows a typical near-field intensity profile of the device at the resonance wavelengths. The long-range variation of peak emitted power across the spectral range is primarily caused by the fixed gap between access waveguide and ring resonator that couples varying power into the resonator across the spectrum.

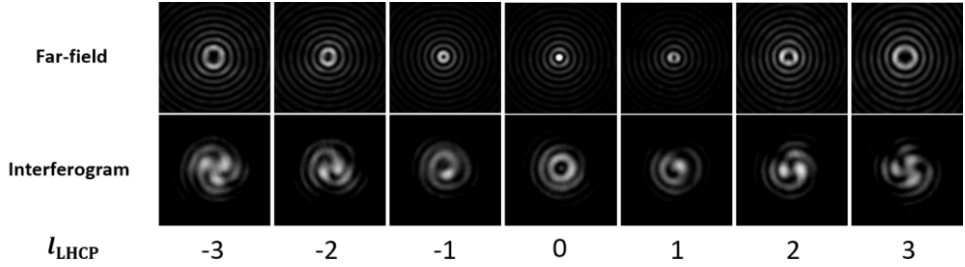

**Supplementary Figure 4.** Far-field profiles and interferograms of left-hand circular-polarized components of CVVs from device $WG_{6-8}$.

Some typical far-field intensity profiles and interferograms of CVVs are illustrated in Supplementary Figure 4, in which the device $WG_{6-8}$ is configured for the emission of CVVs with $l_{TC}$ from 2 to +4. The left-hand circular polarized (LHCP) component is obtained by filtering the far-field CVVs with a QWP and LP combination, and then interferes with the LHCP Gaussian beam. For each CVV of $l_{TC}$, the LHCP component possesses the OAM state of $l_{LHCP} = l_{TC}-1$ (see Equation (4) in the main text), and therefore each interferogram shown in the figure clearly exhibits the spiral fringes with the number of $l_{LHCP}$ [1].

**Supplementary Note 6. Experimental Setup for Spin-Orbit Unidirectional Coupling**

The experimental setup for the measurement of spin-orbit controlled unidirectional coupling is shown in Supplementary Figure 3. The polarized light from the tunable laser is collimated with a collimator and then reflected by the SLM for the conversion to the vortex carrying OAM state $l_{in}$. The linear-polarized vortex is imparted a certain polarization state ($\sigma_{in}$) by the rotatable QWP. A 20X objective lens is used for focusing and illuminating the prepared vortex of spin and orbital AM states $<\sigma_{in}, l_{in}>$ onto the device. Two lensed fibers are used for collecting the received power from the waveguide Ports 1 and 2, respectively.

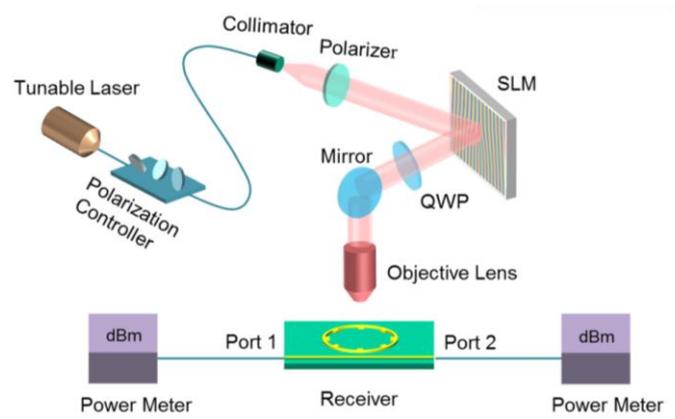

**Supplementary Figure 5.** Experimental setup for the measurement of spin-orbit controlled directional coupling of waveguide modes.

**Supplementary References**

[1] X. Cai, J. Wang, M. J Strain, B. Johnson-Morris, J. Zhu, M. Sorel, J. L. O'Brien, M. G. Thompson, and S. Yu, Science **338**, 363-366 (2012).

[2] J. Zhu, X. Cai, Y. Chen, and S. Yu, Opt. Lett. **38**, 1343-1345 (2013).

[3] J. Zhu, Y. Chen, Y. Zhang, X. Cai, and S. Yu, Opt. Lett. **39**, 4435-4438 (2014).

[4] L. Allen, M. W. Beijersbergen, R. J. C. Spreeuw, and J. P. Woerdman, Phys. Rev. A **45**, 8185-8189 (1992).

[5] S. M. Barnett and L. Allen, Opt. Commun. **110**, 670-678 (1994).

[6] S. Pancharatnam, Proc. Ind. Acad. Sci. **44**, 247-262 (1956).

[7] M. Stalder, and M. Schadt, Opt. Lett. **21**, 1948-1950 (1996).

[8] G. Gibson, J. Courtial, M. J. Padgett, M. Vasnetsov, V. Pas'ko, S. M. Barnett, and S. Franke-Arnold, Opt. Express **12**, 5448-5456 (2004).

[9] M. J. Strain, X. Cai, J. Wang, J. Zhu, D. B. Phillips, L. Chen, M. Lopez-Garcia, J. L. O'Brien, M. G. Thompson, M. Sorel, and S. Yu, Nat. Commun. **5**, 4856 (2014).